\documentclass[sigconf]{acmart}
\usepackage{pifont}
\usepackage{listings}
\usepackage{xcolor}
\usepackage{tikz}
\usepackage{hyperref}
\usepackage{todonotes}
\usepackage{xcolor}
\usetikzlibrary{calc}
\definecolor{softblue}{RGB}{173, 216, 230}
\definecolor{pllmfill}{HTML}{CCD6E3}
\definecolor{smtfill}{HTML}{E1E3EB}
\definecolor{readpyfill}{HTML}{EBEFF2}
\definecolor{centerfill}{HTML}{2874A6}

\usepackage{mdframed}

\newmdenv[
  linewidth=0.5pt,
  roundcorner=2pt,
  innertopmargin=2pt,
  innerbottommargin=2pt,
  innerleftmargin=4pt,
  innerrightmargin=4pt
]{answerbox}

\copyrightyear{2026}
\acmYear{2026}
\setcopyright{cc}
\setcctype{by}
\acmConference[FSE Companion '26]{34th ACM Joint European Software Engineering Conference and Symposium on the Foundations of Software Engineering}{July 05--09, 2026}{Montreal, QC, Canada}
\acmBooktitle{34th ACM Joint European Software Engineering Conference and Symposium on the Foundations of Software Engineering (FSE Companion '26), July 05--09, 2026, Montreal, QC, Canada}
\acmDOI{10.1145/3803437.3808241}
\acmISBN{979-8-4007-2636-1/2026/07}

\begin{document}

\title[SMT-LLM: Constraint-Driven Python Dependency Resolution]{Breaking the Dependency Chaos: A Constraint-Driven Python Dependency Resolution Strategy with Selective LLM Imputation}

\author{Kowshik Chowdhury}
\email{kchowdh1@students.kennesaw.edu}
\affiliation{%
  \institution{Kennesaw State University}
  \city{Marietta}
  \state{Georgia}
  \country{USA}
}

\author{Dipayan Banik}
\email{dipayan5175@gmail.com}
\affiliation{%
  \institution{Danovo Energy Solutions}
  \city{Raleigh}
  \state{North Carolina}
  \country{USA}
}

\author{Shazibul Islam Shamim}
\email{mshamim@kennesaw.edu}
\affiliation{%
  \institution{Kennesaw State University}
  \city{Marietta}
  \state{Georgia}
  \country{USA}
}

\begin{abstract}
Dependency resolution is the task of selecting package versions that can be installed together without conflicts. It accounts for a significant share of build failures in modern software projects. In the Python ecosystem, this task is especially challenging due to Python~2/3 incompatibilities, deprecated packages, and widespread missing metadata. Recent work, such as PLLM, tackles this problem by using large language models (LLMs) to infer Python and package versions from code and iteratively repairing them based on build errors. We present \textsc{SMT-LLM}, a hybrid system that replaces LLM-only version guessing with formal constraint solving. \textsc{SMT-LLM} uses deterministic import extraction and Python version detection via abstract syntax tree (AST) analysis, the \texttt{vermin} tool to infer minimum Python versions, and a five-tier import-to-package resolver that queries PyPI before any LLM call. We construct a constraint graph from PyPI metadata and LLM-imputed dependencies for packages with missing metadata, then solve for consistent version assignments using a Z3 satisfiability modulo theories (SMT) solver. On the HG2.9K benchmark using Gemma2:9B (10\,GB VRAM), \textsc{SMT-LLM} resolves \textbf{83.6\%} of snippets compared to PLLM's 54.8\%, while reducing median resolution time from 151.5\,s to 23.9\,s (\textbf{6.3$\times$ faster}) and average LLM calls from ${\sim}$24.9 to 2.26 per snippet (\textbf{11$\times$ reduction}).
\end{abstract}

\ccsdesc[300]{Software and its engineering~Empirical software validation}
\ccsdesc[500]{Software and its engineering~Software configuration management and version control systems}
\ccsdesc[500]{Software and its engineering~Maintaining software}

\keywords{Python dependency resolution, SMT solving, package version 
conflict, constraint satisfaction, PyPI, LLM-assisted software engineering}

\maketitle
\vspace{-5pt}
\section{Introduction}
\label{sec:intro}

Dependency conflicts and failed dependency resolution are widespread in the Python package ecosystem, regularly breaking builds and delaying development. A study of 235 real-world cases from PyPI projects shows that even small changes to version constraints can introduce new conflicts and installation failures, making dependency management difficult in practice~\cite{watchman}. 
Given only a code snippet—often years old and missing a \texttt{requirements.txt} file—a dependency resolver must choose the right packages, version numbers, and Python interpreter. It must also handle Python 2/3 compatibility issues, deprecated packages, and missing package metadata~\cite{jia2024empirical}. 
Existing LLM-based methods, such as PLLM, use iterative prompting to infer these dependencies, but they still fail on nearly half of the HG2.9K benchmark, leaving many snippets unresolved~\cite{PLLM,dockerizeme}.

\begin{figure*}[!t]
  \centering
  \includegraphics[width=\textwidth]{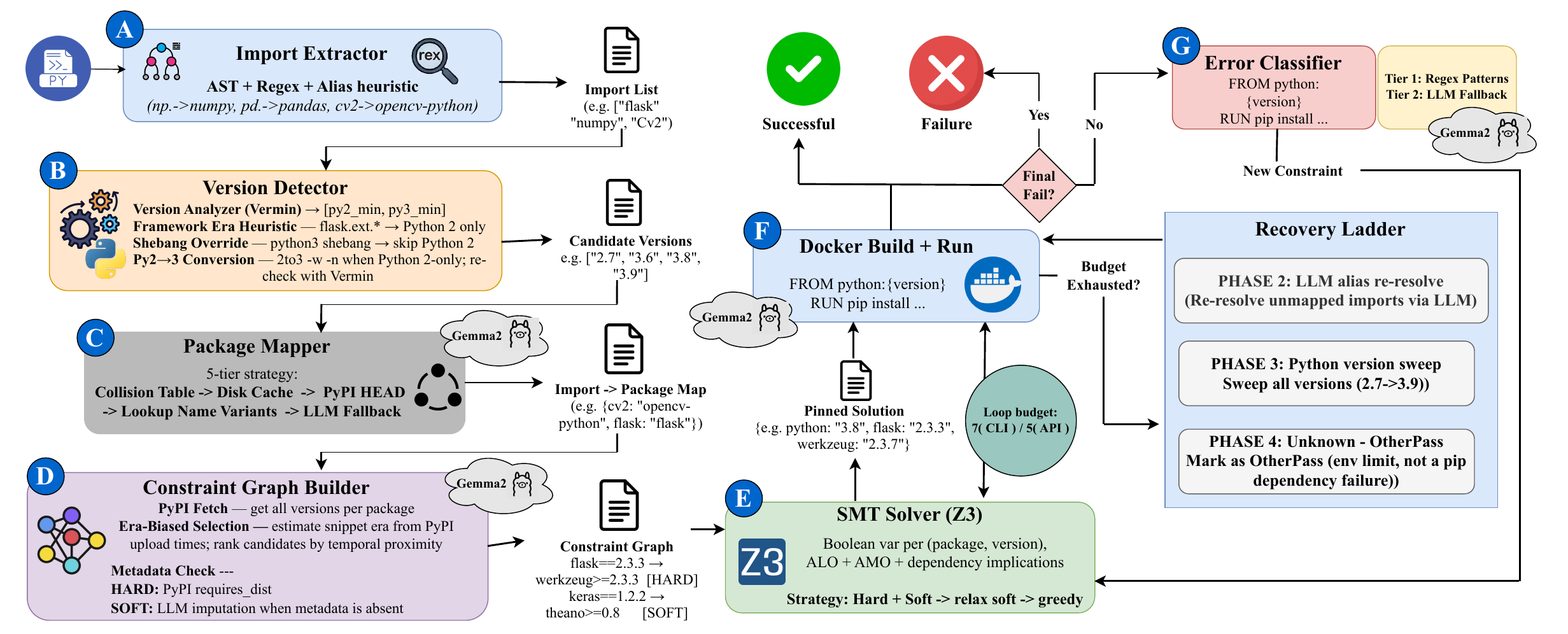}
  \caption{Overview of the \textsc{SMT-LLM} pipeline}
  \label{fig:pipeline}
\end{figure*}

\noindent\textbf{Key Insight.}
Our investigation of PLLM's failures reveals three structural limitations. \textbf{\textit{L1: Initial guess quality.}} PLLM guesses both the Python version and all module versions in a single prompt with no metadata lookup. Versions are chosen independently, so conflicting pairs like \texttt{flask==2.0} (requires \texttt{werkzeug>=2.0}) with \texttt{werkzeug==0.16} go undetected until \texttt{pip install} fails. This also misidentifies the Python version: 17.1\% of snippets produce SyntaxErrors from running Python~2 code under Python~3. \textbf{\textit{L2: Repair loop.}} On failure, PLLM prompts the LLM to replace one module version at a time, ignoring constraint hints pip already provides. If pip reports \texttt{``requires werkzeug>=2.0''}, PLLM discards this and guesses again. This one-at-a-time strategy cannot fix conflicts requiring several packages to change together. \textbf{\textit{L3: Search strategy.}} PLLM launches three Python versions in parallel but never shares outcomes across them. Python~3.6 may fail five times while 3.8 has already succeeded, yet 3.6 keeps retrying because no early-termination signal propagates across parallel tracks.

Our \textsc{SMT-LLM} approach addresses these limitations by restricting the LLM to verifiable factual queries (e.g., ``What are \texttt{flask==
0.10}'s \texttt{install\_requires}?'') and delegating everything else to deterministic tools. Specifically, SMT-LLM (i)~replaces version guessing with AST-based analysis and a five-tier PyPI resolver (addressing L1), (ii)~parses \texttt{pip} errors into formal Z3~\cite{z3} constraints instead of blind re-prompting (addressing L2), and (iii)~tests Python versions sequentially, stopping at the first success (addressing L3). The contributions are: \noindent\ding{182}~A \textit{seven-stage hybrid pipeline} (Figure~\ref{fig:pipeline}) combining AST-based static analysis, a five-tier PyPI resolver, LLM-imputed constraint graphs, and Z3 SMT solving with Docker-based validation (\S\ref{sec:approach}).
\ding{183}~A \textit{hard/soft constraint distinction} enabling the solver to trust PyPI metadata while treating LLM-imputed dependencies as relaxable (\S\ref{sec:approach}, Stage~D).
\ding{184}~An \textit{error-driven constraint refinement loop} that parses Docker failures into formal Z3 constraints, strictly narrowing the search space (\S\ref{sec:approach}, Stage~G).
\ding{185}~A \textit{Recovery Ladder} with LLM alias re-resolution, deterministic Python version sweep, and environment-limitation classification (\S\ref{sec:approach}, Stage~G).
\ding{186}~Evaluation on HG2.9K showing an 83.6\% resolution rate, with substantially fewer LLM calls per snippet than PLLM (2.26 vs. ${\sim}$24.9) (Table~\ref{tab:efficiency}).

\vspace{-5pt}
\section{The SMT-LLM Pipeline}
\label{sec:approach}

\sloppy

We retain PLLM's Docker-based validation backend but replace its LLM-centric pipeline with seven stages of deterministic analysis, Z3 constraint solving, and selective LLM imputation (Gemma-2:9B~\cite{gemma2} via Ollama, 10\,GB VRAM), as illustrated in Figure~\ref{fig:pipeline}.

\par\vspace{1pt}\noindent\textbf{Stage A: Import Extraction.} SMT-LLM parses \texttt{Import} and \texttt{ImportFrom} nodes via AST, falling back to regex for Python~2 files that fail parsing~\cite{minyi_ast_deep_dive}. Standard-library modules are filtered using \texttt{sys.stdlib\_module\_names}, a hardcoded Python~2 stdlib set, and runtime \texttt{importlib.util.find\_spec()}. Django-style submodule imports (e.g., \texttt{from X.models import Y}) are excluded as project-local code. Snippets whose imports map exclusively to non-Linux, platform-specific modules (Sublime Text APIs, macOS Cocoa bindings, Blender internals, \texttt{winreg}, \texttt{win32api}) are labeled \texttt{OtherPass} and bypass Docker entirely. These modules are embedded in their host applications, not distributed via PyPI, so detecting them early prevents unnecessary Docker builds that cannot succeed regardless of the selected version.

\par\vspace{1pt}\noindent\textbf{Stage B: Python Version Detection.} We use \texttt{vermin}\footnote{\url{https://github.com/netromdk/vermin/}}, a static analysis tool that inspects AST node types to infer the minimum compatible Python~2 and Python~3 versions. When \texttt{vermin} cannot determine compatibility, the pipeline falls back to \texttt{[2.7,~3.6,~3.8,~3.9]}. For Python~2-only snippets, the pipeline invokes \texttt{2to3} within a \texttt{python:3.9} Docker container when the local \texttt{lib2to3} module is unavailable (removed in Python~3.13+). This avoids deprecated Python~2.7 Debian Buster images whose EOL apt mirrors are increasingly unreliable.

\par\vspace{1pt}\noindent\textbf{Stage C: Import-to-Package Mapping.} Each import must resolve to its PyPI distribution (e.g., \texttt{sklearn}$\rightarrow$\texttt{scikit-learn}). The collision table contains 36~curated mappings from PyPI metadata and mapping databases (e.g., pipreqs~\cite{pipreqs}). Tier~3 sends parallel PyPI HEAD requests testing exact-case, lowercase, and capitalized variants. Tier~4 applies nine structural name-variant patterns (e.g., \texttt{python-\{name\}}, \texttt{py\{name\}}). The LLM fallback (Tier~5) is accepted only if validated against PyPI. The pipeline persists all non-trivial mappings to disk for reuse in subsequent runs.

\par\vspace{1pt}\noindent\textbf{Stage D: Constraint Graph Construction.} For each package, the PyPI JSON API enumerates candidate versions after filtering yanked releases and Python-incompatible builds. Wheel-available versions are sorted first to avoid source-compilation failures on legacy interpreters. The two-pass era-biased selection picks up to eight candidates: the first pass estimates the snippet's authorship era as the median of per-package midpoint PyPI upload times; the second re-ranks by temporal proximity, keeping the five nearest plus three uniformly sampled from the remainder. This limit keeps Z3's Boolean variable count in the millisecond regime; the Docker retry loop covers versions outside this window. \texttt{requires\_dist} metadata (PEP~508 specifiers, e.g., \texttt{werkzeug>=2.3.3}) produces \textbf{hard} edges. When \texttt{requires\_dist} is null, common in pre-2015 packages -- the LLM imputes dependencies as a factual recall task (e.g., \textit{``What are the direct pip-install dependencies of \texttt{theano==0.9.0}?''}), producing \textbf{soft} edges (relaxable on UNSAT). Packages with zero installable versions are replaced before the Docker loop begins.

\par\vspace{1pt}\noindent\textbf{Stage E: SMT Solving.}
The constraint graph is encoded as a Z3~\cite{z3} Boolean satisfiability instance, with a Boolean variable for each (package, version) pair. The constraint classes are asserted: \textbf{(1)}~\emph{At-Least-One} (ALO): every package in the import list must have at least one version selected, ensuring no dependency is left unresolved; \textbf{(2)}~\emph{At-Most-One} (AMO): each package can have at most one version selected, preventing conflicting installs of the same package. Together, ALO and AMO guarantee exactly one version 
per package; \textbf{(3)}~dependency implications ensure that selecting a version entails selecting a compatible version of each declared dependency, with package names normalized across hyphens, underscores, and dots. The solver first attempts satisfiability with all hard and soft constraints; on UNSAT, it retries with soft constraints relaxed; if still UNSAT, a greedy fallback selects the second-newest version per package. The solver terminates in under one second, yielding a globally consistent pinned environment.

\par\vspace{1pt}\noindent\textbf{Stage F: Docker Validation.}
The Dockerfile targets \texttt{{-}{-}platform=linux/amd64}. For EOL Buster images (Python~2.7, 3.6, 3.7), apt sources are redirected to \texttt{archive.debian.org}. Before the first build, solved packages are checked against a curated table of~46 C-extension packages and their Debian build dependencies (e.g., \texttt{scipy}$\rightarrow$\texttt{gfortran+libopenblas-dev}); matching apt packages are injected into the Dockerfile, eliminating a wasted first-build failure for known native-code packages. Packages install via BuildKit pip-cache mounts shared across builds. Build and run timeouts are 450\,s and 60\,s; a clean exit denotes \texttt{Pass}. When validation fails, the pipeline enters the error classification and re-solving stage.

\par\vspace{1pt}\noindent\textbf{Stage G: Error Classification and Re-Solving.}
Docker failures are classified via an ordered, first-match regex taxonomy with eleven types: \textit{VersionNotFound}, \textit{DependencyConflict}, \textit{ModuleNotFound}, \textit{ImportError}, \textit{SyntaxError}, \textit{NonZeroCode}, \textit{AttributeError}, \textit{SystemLibError}, \textit{ContainerTimeout}, \textit{EnvironmentErrorFallback}, and \textit{ExecutionError}; an LLM fallback handles unmatched logs. Each error injects a new constraint; Z3 re-solves and Docker retries for up to five iterations, with a deduplication guard on repeated (python, packages, apt) states. \texttt{NonZeroCode} consults the apt build-dependency table and attempts a binary-variant swap (e.g., \texttt{psycopg2}$\rightarrow$\texttt{psycopg2-binary}) before any LLM call. The Recovery Ladder's version sweep uses a three-iteration budget per candidate.

\fussy

\vspace{-5pt}
\section{Results}
\label{sec:result}
\begin{figure}[H]
\centering
\begin{tikzpicture}[font=\small, scale=0.6]
\def\r{2.25}
\coordinate (A) at (0,0);
\coordinate (B) at (2.8,0);

\fill[pllmfill] (A) circle (\r);
\fill[smtfill] (B) circle (\r);

\begin{scope}
  \clip (A) circle (\r);
  \fill[centerfill] (B) circle (\r);
\end{scope}

\draw[line width=1.2pt, black!80] (A) circle (\r);
\draw[line width=1.2pt, black!80] (B) circle (\r);

\node[above] at ($(A)+(0,\r+0.1)$) {PLLM};
\node[above] at ($(B)+(0,\r+0.1)$) {SMT-LLM};

\node[align=center] at ($(A)+(-1.0,0)$) {66\\(3\%)};
\node[align=center, font=\bfseries, text=white] at ($(A)+(1.4,0)$) {1517\\(61\%)};
\node[align=center] at ($(B)+(1.0,0)$) {900\\(36\%)};
\end{tikzpicture}
\caption{\small Overlap of successfully fixed snippets: PLLM vs SMT-LLM}
\label{fig:venn-fix-overlap}
\end{figure}
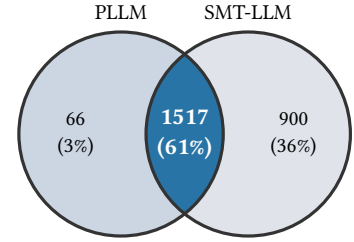

We evaluate SMT-LLM with PLLM, the strongest baseline by fix rate, on 2,891 benchmark gists \cite{PLLM}. Table \ref{tab:efficiency} summarizes the overall performance and Figure \ref{fig:venn-fix-overlap} covers the 2,483 snippets resolved by at least one tool; the remaining 408 (14.1\%) could not be fixed by either tool. The failed snippets are dominated by dead or renamed PyPI packages that neither tool can map to an installable dependency.
\begin{table}[H]
\centering
\caption{\small Efficiency comparison: PLLM vs SMT-LLM}
\label{tab:efficiency}
\small
\setlength{\tabcolsep}{4pt}
\begin{tabular}{@{}lrr@{}}
\toprule
& \textbf{PLLM} & \textbf{SMT-LLM} \\
\midrule
Success Rate              & 54.8\%       & 83.6\%  \\
Median Time (s)           & 151.5        & 23.9    \\
P90 Time (s)              & 491.0        & 186.6   \\
\midrule
\multicolumn{3}{@{}l}{\textit{Version Detection}} \\
SyntaxError Rate          & 17.1\%       & 0.5\%   \\
No-LLM Pass              & 0\%          & 45.0\%  \\
First-Build Pass          & ${\sim}$8\%  & 42.5\%  \\
\midrule
\multicolumn{3}{@{}l}{\textit{Repair Loop}} \\
LLM Calls / Snippet       & ${\sim}$24.9 & 2.26    \\
Docker Iters / Snippet     & ${\sim}$23.9 & 4.9     \\
\midrule
\multicolumn{3}{@{}l}{\textit{Search Strategy}} \\
Single-Version Pass       & 0\%          & 62.5\%  \\
\bottomrule
\end{tabular}
\end{table}

SMT-LLM fixes 2,417 snippets (83.6\%), a 28.8 percentage-point improvement over PLLM's 54.8\%. Of these, 1,517 are shared fixes, indicating a common core of easy repairs, while SMT-LLM uniquely resolves 900 snippets compared to only 66 by PLLM. Beyond fix rate, SMT-LLM reduces median resolution time from 151.5\,s to 23.9\,s (6.3$\times$ speedup), driven by three architectural advances. First, AST-based Python version detection reduces SyntaxError rates from 17.1\% to 0.5\%; combined with constraint-aware package selection, 42.5\% of first Docker builds succeed without retry and 45\% of successful resolutions require zero LLM calls. Second, structured error feedback replaces PLLM's re-prompting: pip error messages are parsed into Z3 constraints, reducing average LLM calls from ${\sim}$24.9 to 2.26 and Docker iterations from 23.9 to 4.9 per snippet. Third, sequential version testing with early termination resolves 62.5\% of snippets on the first candidate version alone. This eliminates PLLM's redundant parallel computation, where all three versions run to completion even after one succeeds. LLM stochasticity has limited impact: calls use temperature~0.1 with local caching, 45\% of resolutions require zero LLM calls, and the primary non-determinism source is Z3, not the LLM (Section~\ref{sec:limitations}).

\vspace{5pt}

\begin{answerbox}
\textbf{Key Result.} Compared to PLLM, SMT-LLM improves the fix rate from 54.8\% to \textbf{83.6\%}, reduces median resolution time by \textbf{6.3×}, reduces LLM calls by \textbf{11×}, and requires \textbf{5×} fewer Docker iterations, with \textbf{45\%} of successful resolutions without any LLM calls.  
\end{answerbox}

\vspace{2pt}

\vspace{-5pt}

\section{Limitations and Root-Cause Analysis}
\label{sec:limitations}

Table~\ref{tab:failure-breakdown} summarizes the 474 failures by root cause. Missing modules dominate, from platform-embedded SDKs (Pythonista, IDA~Pro, Rhino3D, Maya) and project-local imports absent from PyPI. Build /wheel failures trace largely to Python\,2-only bindings lacking Python\,3 releases
(\texttt{pygtk} alone accounts for 64\% of VersionNotFound cases). Import and attribute errors reflect restructured internals (e.g.,\ App~Engine NDB migration, removed IPython and \texttt{ggplot} APIs). The remaining cases
comprise Python\,2 syntax (\texttt{print} statements, \texttt{ur''} prefixes) resisting \texttt{2to3}, and Blender \texttt{bpy} scripts requiring an embedded runtime unreachable through pip. Additionally, both PLLM and SMT-LLM exhibit non-deterministic behavior in version selection: PLLM because the LLM's predictions vary across invocations,
and SMT-LLM because the Z3 solver may return different valid assignments when multiple satisfying solutions exist.

\vspace{-5pt}

\section{Threats to Validity}

\textbf{Local-module ambiguity.}
Static import extraction cannot distinguish project-local modules from genuine PyPI packages.  When a gist snippet is extracted from a larger project, imports such as \texttt{from protobuf import IpcConnectionContext\_pb2} (gist \texttt{5441636}---a
local directory of generated Protocol Buffer files), \texttt{import settings} (gist \texttt{4413028}---a Django project settings file) are all treated as third-party dependencies.  SMT-LLM exhausts its replacement heuristics before dropping the unresolvable import, wasting Docker iterations. We partially mitigate this by filtering identity-mismatch drops from the final module list, but local modules that fail to build remain indistinguishable from genuinely deprecated packages.

\textbf{PyPI temporal drift.}
The HG2.9K gists were authored between 2011 and 2019, but the constraint graph queries today's PyPI index. Packages have since been yanked, renamed, or stripped of older wheels, creating a mismatch between what snippets originally required and what remains available. Despite era-biased version selection, this accounts for 70 of 474 failures (14.8\%): \emph{VersionNotFound}~(36) when no compatible wheel exists for the target Python, and \emph{NonZeroCode}~(34) when source distributions fail to compile against modern system libraries. Pinning to a historical PyPI snapshot could reduce this gap but is beyond our current scope.

\textbf{Cache accumulation.}
The persistent mapping cache accelerates resolution by reusing previously verified import-to-package mappings across runs.  However, an incorrect LLM-generated mapping (e.g., \texttt{block\_diag}~$\to$~\texttt{blockdiag} instead of recognizing it as \texttt{scipy.linalg.block\_diag}) becomes a cached false mapping that silently affects all subsequent snippets sharing that import.  We address this through periodic manual audits of the cache, but systematic validation against PyPI metadata remains future work.

\begin{table}[!t]
\centering
\caption{\small Root-cause breakdown of 474 SMT-LLM failures}
\label{tab:failure-breakdown}
\small
\setlength{\tabcolsep}{4pt}
\begin{tabular}{@{}llc@{}}
\toprule
\textbf{Category} & \textbf{Root Cause} & \textbf{N (\%)} \\
\midrule
Missing modules   & Platform SDKs, local imports     & 276 (58.2) \\
Import errors  & Deprecated or renamed internals  & 65 (13.7)  \\
Version not found   & Needed version not on PyPI       & 36 (7.6)   \\
Non-zero exit   & Runtime failure after install     & 34 (7.2)   \\
Attribute errors   & API changes across versions       & 34 (7.2)   \\
Syntax errors  & Python\,2 syntax resists 2to3    & 15 (3.2)   \\
System libraries      & Missing OS libs (GTK, RPi)       & 14 (3.0)   \\
\hline
\end{tabular}
\end{table}

\vspace{-5pt}
\section{Related Work}
Python dependency resolution has attracted growing attention as the ecosystem's package volume and version-conflict frequency continue to rise~\cite{jia2024dependency}. Horton and Parnin~\cite{horton2018gistable} introduced the HG2.9K benchmark of GitHub gists, exposing the brittleness of standard tools such as \texttt{pip} and \texttt{pipreqs}, which rely on greedy backtracking and lack cross-package global reasoning. In the broader software engineering landscape, SAT and SMT solvers have long underpinned package managers for other ecosystems; Debian's \texttt{apt} models upgrades as a pseudo-Boolean optimization problem~\cite{mancinelli2006managing}, and Eclipse's p2 provisioning uses SAT to resolve OSGi bundles~\cite{leberre2009dependency}, yet these techniques have seen limited adoption in Python's loosely specified, metadata-sparse packaging ecosystem. Concurrently, LLMs have demonstrated effectiveness in code repair~\cite{xia2023aprllm}, test generation~\cite 
{lemieux2023codamosa}, and fault localization~\cite{yang2024llmao}, but their use as \emph{constraint generators} rather than end-to-end solvers remains underexplored. SMT-LLM bridges this gap by using the LLM only to fill metadata gaps and classify errors, while delegating version selection to Z3.

\vspace{-5pt}

\section{Conclusion}

\textsc{SMT-LLM} resolves 2,417 of 2,891 HG2.9K snippets (83.6\%), combining Z3 constraint solving with selective LLM imputation to replace PLLM's iterative guess-and-check loop. The 474 remaining failures stem from platform-specific SDKs (Sublime Text, IDA Pro, Blender), project-local modules absent from PyPI, and legacy C-extension builds whose binary wheels no longer exist; the practical ceiling for container-based resolution against a live package index. This represents a 52.6\% relative improvement over PLLM's 54.8\% while requiring $11\times$ fewer LLM calls and $5\times$ fewer Docker iterations per snippet. These results suggest that when a problem splits into factual lookup and combinatorial search, restricting the LLM to verifiable queries and delegating version selection to a formal solver like Z3 is faster, cheaper, and more reproducible than end-to-end neural reasoning.

\noindent\textbf{Future work.} We plan to address the Z3 solver
non-determinism through deterministic pinning strategies, integrate
historical PyPI snapshots to reduce temporal-drift failures, and develop
a classifier to distinguish project-internal imports from third-party
dependencies before resolution begins.

\vspace{-5pt}

\section{Data Availability}
To support reproducibility, the SMT-LLM implementation, Docker runner, and all evaluation artifacts are publicly available~\cite{smtllm2025repo}.

\vspace{-5pt}
\bibliographystyle{ACM-Reference-Format}
\bibliography{references}

@article{PLLM,
  title={The Last Dependency Crusade: Solving Python Dependency 
         Conflicts with LLMs},
  author={Antony Bartlett and Cynthia Liem and Annibale Panichella},
  journal={2025 40th IEEE/ACM International Conference on Automated 
           Software Engineering Workshops (ASEW)},
  year={2025},
  pages={66--73},
  url={https://api.semanticscholar.org/CorpusID:275921543}
}

@misc{gemma2,
  author       = {Ollama},
  title        = {Gemma 2 9B},
  year         = {2026},
  howpublished = {\url{https://ollama.com/library/gemma2:9b}},
  note         = {[Online; accessed 05-March-2026]}
}

@misc{minyi_ast_deep_dive,
  author       = {Min Yi},
  title        = {Abstract Syntax Tree (AST) Deep Dive: From Theory to Practical Compiler Implementation},
  year         = {2026},
  howpublished = {\url{https://dev.to/min_yi_e5fbf986e24f1c42df/abstract-syntax-tree-ast-deep-dive-from-theory-to-practical-compiler-implementation-4jpo}},
  note         = {[Online; accessed 05-March-2026]}
}

@inproceedings{jia2024empirical,
  title={An empirical study on Python library dependency and conflict issues},
  author={Jia, Xinyu and Zhou, Yu and Hussain, Yasir and Yang, Wenhua},
  booktitle={2024 IEEE 24th International Conference on Software Quality, Reliability and Security (QRS)},
  pages={504--515},
  year={2024},
  organization={IEEE}
}

@inproceedings{ z3,
  title={Z3: An efficient SMT solver},
  author={De Moura, Leonardo and Bj{\o}rner, Nikolaj},
  booktitle={International conference on Tools and Algorithms for the Construction and Analysis of Systems},
  pages={337--340},
  year={2008},
  organization={Springer}
}

@misc{pipreqs,
  author = {V. Kravcenko},
  title  = {pipreqs: Generate \texttt{pip} \texttt{requirements.txt} based on imports},
  year   = {2023},
  howpublished = {\url{https://github.com/bndr/pipreqs}}
}

@inproceedings{dockerizeme,
  title={Dockerizeme: Automatic inference of environment dependencies for python code snippets},
  author={Horton, Eric and Parnin, Chris},
  booktitle={2019 IEEE/ACM 41st International Conference on Software Engineering (ICSE)},
  pages={328--338},
  year={2019},
  organization={IEEE}
}

@article{jia2024dependency,
  author    = {Jia, Zhijie and others},
  title     = {A Survey on Python Dependency Conflicts},
  journal   = {IEEE Transactions on Software Engineering},
  year      = {2024},
  doi       = {10.1109/TSE.2024.3458529},
}

@inproceedings{horton2018gistable,
  author    = {Eric Horton and Chris Parnin},
  title     = {Gistable: Evaluating the Executability of Python Code Snippets on {GitHub}},
  booktitle = {Proc. IEEE International Conference on Software Maintenance and Evolution (ICSME)},
  year      = {2018},
  doi       = {10.1109/ICSME.2018.00029},
}

@inproceedings{mancinelli2006managing,
  author    = {Fabio Mancinelli and Jaap Boender and Roberto Di Cosmo and Jerome Vouillon},
  title     = {Managing the Complexity of Large Free and Open Source Package-Based Software Distributions},
  booktitle = {Proc. IEEE/ACM International Conference on Automated Software Engineering (ASE)},
  year      = {2006},
  doi       = {10.1109/ASE.2006.49},
}

@inproceedings{leberre2009dependency,
  author    = {Daniel {Le Berre} and Pascal Rapicault},
  title     = {Dependency Management for the Eclipse Ecosystem},
  booktitle = {Proc. International Workshop on Open Component Ecosystems (IWOCE)},
  year      = {2009},
  doi       = {10.1145/1595800.1595803},
}

@article{xia2023aprllm,
  author    = {Chunqiu Steven Xia and Yuxiang Wei and Lingming Zhang},
  title     = {Automated Program Repair in the Era of Large Pre-Trained Language Models},
  journal   = {Proc. IEEE/ACM International Conference on Software Engineering (ICSE)},
  year      = {2023},
  doi       = {10.1109/ICSE48619.2023.00129},
}

@inproceedings{lemieux2023codamosa,
  author    = {Caroline Lemieux and Jeevana Priya Inala and Shuvendu K. Lahiri and Siddhartha Sen},
  title     = {{CodaMosa}: Escaping Coverage Plateaus in Test Generation with Pre-Trained Large Language Models},
  booktitle = {Proc. IEEE/ACM International Conference on Software Engineering (ICSE)},
  year      = {2023},
  doi       = {10.1109/ICSE48619.2023.00085},
}

@inproceedings{yang2024llmao,
  author    = {Aidan Z.H. Yang and Claire {Le Goues} and Ruben Martins and Vincent J. Hellendoorn},
  title     = {Large Language Models for Test-Free Fault Localization},
  booktitle = {Proc. IEEE/ACM International Conference on Software Engineering (ICSE)},
  year      = {2024},
  doi       = {10.1145/3597503.3623342},
}

@misc{smtllm2025repo,
  author       = {Chowdhury, Kowshik},
  title        = {A hybrid SMT + selective-LLM pipeline for Python dependency resolution (For FSE-AIWare ’26)},
  year         = {2026},
  howpublished = {\url{https://github.com/Kowshik-18/SMT-LLM}},
  note         = {[Online; accessed 10-March-2026]}
}

@inproceedings{watchman,
  title={Watchman: Monitoring dependency conflicts for python library ecosystem},
  author={Wang, Ying and Wen, Ming and Liu, Yepang and Wang, Yibo and Li, Zhenming and Wang, Chao and Yu, Hai and Cheung, Shing-Chi and Xu, Chang and Zhu, Zhiliang},
  booktitle={Proceedings of the ACM/IEEE 42nd international conference on software engineering},
  pages={125--135},
  year={2020}
}

\end{document}